# Towards an Understanding of Valence in E-Government Services

**Jason Simpson**
School of Business, Government & Law
University of Canberra
Canberra, Australian Capital Territory
Email: jason.simpson@canberra.edu.au

**John Campbell**
School of Business, Government & Law
University of Canberra
Canberra, Australian Capital Territory
Email: john.campbell@canberra.edu.au

**Thuy Pham**
School of Business, Government & Law
University of Canberra
Canberra, Australian Capital Territory
Email: thuy.pham@canberra.edu.au

**Byron Keating**
Research School of Management
The Australian National University
Canberra, Australian Capital Territory
Email: byron.keating@anu.edu.au

**Carla Wilkin**
School of Accounting
Monash University
Melbourne, Victoria
Email: carla.wilkin@monash.edu

## Abstract

The Australian government, to remind job seekers of appointments with employment services providers in order to cut costs and free up human resources, is using technologies such as Short Messaging Services (SMS). However, the technologies in-use are but one side of this equation – the specifics of how these technologies are used is the other side, and these specifics are highly under-theorized, particularly in regard to the views of the people to which these technologies are directed. The purpose of this paper is to provide a theoretical framing for this phenomenon as well as to introduce an emerging methodological direction that may allow for a better understanding of demographic-specific values and thereby better valence framing. The paper also theorizes reactions to information that could be applicable elsewhere, not just in e-government or with SMS, thereby contributing to discussions surrounding the 'Big Data' debate

## 1    Introduction

Bélanger and Carter (2012) define e-government as "the use of information technology to enable and improve the efficiency with which government services are provided to citizens, employees, businesses and agencies" (p. 364). This paper argues that understanding valence in general, and then later specifically with respect to Short Messaging Service (SMS) texts (for example) directed at the unemployed/underemployed, might help governments reduce costs, as well as generate public value (cf. Pang et al. 2014). The use of such technologies is only one side of saving money / freeing up resources. The other side is the specifics of how these technologies are used, which is under-theorized particularly in regard to the views of the people to which these technologies are directed. Prospect theory (Kahneman and Tversky 1979) offers a way of viewing decision making in terms of gain/loss, positive/negative, which provides a top-down approach to framing SMS valence, and Personal Construct Theory (Kelly 1955) offers a bottom-up approach to *understanding* what the unemployed/underemployed actually consider to be gain/loss, positive/negative. Combined, the two offer an approach for engaged scholarship (Van de Ven 2007) in e-government, particularly when informed by both pragmatism (Dewey 1929) and relationality.

IS scholars have studied a variety of e-government topics such "e-reverse auctions, GIS, online license renewal, mobile data services, and so forth" (Bélanger and Carter 2012, p. 377) using a variety of methods and perspectives (ibid). However, e-government studies in top IS journals or highly cited articles have yet to explore in-depth what constitutes value in the eyes of citizens (ibid). In the case of the Australian government, much financial support is directed through income support for the unemployed/ underemployed. While the government has taken steps to address the situation by requiring those on income support to regularly attend appointments with employment providers, some unemployed/underemployed job seekers fail to show up for scheduled appointments even under threat of losing financial benefits. A simple explanation might be that the job seeker simply forgot; however, SMS reminders of scheduled appointments sent the day before each appointment have had mixed impact.

To address this problem and answer calls for research into value in the eyes of citizens, particularly those outside of the US (Bélanger and Carter 2012), pilot research was conducted to discover, at a basic level, what concerns Australian job seekers have and what they consider important with respect to employment service providers. A thematic analysis of the data revealed that the way that information is communicated and/or the way that providers interact with seekers in person is key to understanding valence. Furthermore, the most commonly expressed concern was failure of the provider to understand the individual's circumstances (i.e. not listening). Indeed, job seekers comprise a very specific demographic – one where views on topics surrounding authority, motivations, access to information and technology, and so on may be totally alien to those who do not fall into this particular demographic, especially those in government or the providers. The tendency is to simply ignore these views. However, if one subscribes to a relational view (and we do), problems of this sort are co-constituted and co-created both in their current form as well as their antecedents. As such, we believe that as we all are attempting to navigate the information frontier in a networked society, research into the views of all 'sides' and how they process and understand information increasingly becomes a practical *necessity* rather than something interesting to ponder.

The purpose of this paper is twofold: 1) To provide a theoretical framing that explains why the job seeker demographic may choose to not show up to their appointments as well as why the framing is crucial due to the nature of SMS (it is one way/there is no room for discussion, and the number of words are very limited) and, 2) To provide an emerging methodological direction for future research regarding job seekers' motivations for showing up to their appointments, as well as how technologies in-use might fit into these motivations, in order to increase the effectiveness of digital communication in e-government. We also suggest a similar approach with respect to the providers and the government in order to increase understanding of multi-level stakeholder alignment (Campbell et al. 2013).

## 2 E-Government and Initial Findings

### 2.1 Identifying Opportunities for Research

Bélanger and Carter's (2012) historical review on e-government illustrates the many e-government studies that have been published in top IS journals as well as studies that are highly cited (as well as why these two groupings are not the same). They point out that "e-government is receiving increased attention from researchers and practitioners alike" (p. 366) as there have recently been several special issues in our top IS journals dedicated to the topic (ibid). The purpose of their review was to synthesize findings, identify issues, and provide recommendations for future research (ibid). They also highlight how many early e-government studies aimed at practice were highly a-theoretical but grounded as they attempted to hash out the concept, while top IS journals identified at least one specific theory in-use but were more focused on a specific topic and often (over half the time) appropriated fashionable IS theories rather than considering those outside of the discipline (ibid). Our paper considers both as well as the various calls for research by Bélanger and Carter (2012) in order to bridge this gap between theory and practice, and goes one step beyond the identified gaps for future research. We do so not just because there has been considerable debate around the topic of bridging this type of gap (e.g. Straub and Ang 2011) but also because we hold a relational view/ontology in which neither of these things exists on its own, even if one chooses to *focus* on one or the other (cf. knowledge vs. experience). The reader should keep this notion of relationality salient as it will become increasingly important throughout this paper.

In addition to a theoretical presence comparison between top IS journals and articles that are highly cited, Bélanger and Carter (2012) extract methodological approaches, sampling, and topic areas among the two sets of literature. Methodologically, there were no articles in their review that were conceptual at the individual level of analysis, and this was the most under-researched combination in either set of literature. With respect to sampling, the "majority of the early "highly-cited" e-government articles sample US citizens" (p. 376), and the authors call for research using non-US sampling. Finally, with respect to topic areas, they found that leading IS journals tend to focus on specific services whereas highly cited articles do not (ibid). However, the key point was that perhaps we should consider more fundamental questions such as what "is the value of e-government to citizens and agencies?" and in doing so "researchers should determine what constitutes e-government success or failure from both the government's and citizen's perspectives. In doing so, researchers can help inform practice by helping agencies avoid failure" (p. 378). Interestingly enough, our current research at the Australian Department of Employment stands to address each of these gaps and calls for research – the context of which is described next.

## 2.2  Research Context – Employment Services Australia

In order to address the aforementioned practical concerns, we describe the practical problems faced in our project in order to 1) provide the context for which the early research was carried out, 2) why the context lends to the aforementioned research gaps, and 3) set the stage for the forthcoming theorizing and future research sections.

In Australia, the Department of Employment works to provide job seekers access to services and support to gain employment in partnership with contracted service providers who are engaged to deliver a range of targeted programs. The largest of these programs is Job Services Australia (JS), which provides social welfare recipients with access to a national network of employment support services. As part of a larger research program, we are working in conjunction with Job Services Australia in the Department of Employment to help solve the problem of job seekers not showing up for appointments with employment service providers. The problem is significant, as in the last year around 30 million appointments were made yet only around 8.5 million of those appointments resulted in the job seeker showing for the appointment. This results in providers being unable to meet their targets for employment placements and, consequently, reduces the likelihood of job seekers finding appropriate employment.

At first glance, the situation is confusing, as the government is providing a free service to help those without work to find and secure work, and the appointment with the provider has been made with the job seeker – why would they then not show? There are several possible hasty rationales. First and most obvious, maybe the job seeker simply does not want to work and continue to receive income support. However, steps have been taken where if a job seeker fails to show after a certain number of times they are financially penalized and/or completely cut off from financial support – yet many still do not to show. Second, perhaps the seeker simply forgot. However, as the appointment approaches, the delivery of an SMS message to the job seeker's phone is triggered in order to remind them of their appointment – and yet still, many do not show.

This practical problem provides an opportunity to further theorize the underlying psychology at a more fundamental level by first simply asking the job seekers (individuals) what was important to them (values) with respect to providers (services), and then looking for theories that might help to explain the results (conceptual development) as well as be applicable to SMS messages regarding the government led service (e-government). Taken together, this would begin to contribute to the aforementioned calls for research into individual level conceptual analyses of value in the eyes of citizens with respect to e-government. The results would provide input into the second iteration of (primary) data collection being proposed – the results of which could then be applied to all forms of communication, not just SMS, as the underlying psychology behind e-government related communication would be addressed in some meaningful way.

## 2.3  Pilot Study and Initial Findings

A pilot study was conducted using a text based on-line survey in order to establish some understanding of what is important in the mind of job seekers (i.e. what do citizens in our particular service context consider to be of value, or the opposite of value). Job seekers were randomly selected and asked to participate in the survey. The job seekers were asked to identify what they considered to be the three most important attributes (respectively) that would describe a good employment service. The participants were then asked to explain what was meant by each attribute and why it was important. In total, there were 65 completed surveys.

The data was coded and thematically analysed in conjunction with Dedoose software in order to determine the *most* important concerns across the group of the job seekers regarding providers' employment services. Table 1 lists the most prominent categories and qualifying sub-categories based on code presence. Statements, along with respective codes, naturally fell into positive statements (e.g. "helps with updating resumes and doing interviews") or negative statements (e.g. "I am treated as a number instead of an individual"). The purpose of coding negative statements (statements that do not address features of a good employment service) is to remain in line with the relational view as well as to gain an understanding of what something is by understanding what it is not. There were a total of 362 excerpts, 143 codes, and 949 code applications. Subordinate categories constitute superordinate categories, i.e. recognition of "Fit / suitability of potential position" is one example of how job seekers feel that their "Individual circumstances" have been understood.

|          | Superordinate Category | Subordinate Categories |
|----------|------------------------|-------------------------|
| Positive | Individual circumstances understood (Value) | Fit / suitability of potential position; Understanding of what seeker knows; Understanding seeker's concerns; Recognition of financial, mental, age, educational, etc. limitations; Sensitivity to personal time requirements (e.g. time needed for children) |
|          | Helpful (Chosen attribute) | Honesty; Willingness to help; Finding / Locating jobs; Help with resume; Teach job search / secure processes |
| Negative | Individual circumstances ignored (Value) | Treated as a number / Lost in the crowd; Disinterest in individual; Provider tuning out / Not listening; There to collect a pay check / meet numbers; Failure to recognize financial, mental, age, educational, etc. limitations |

*Table 1. Seekers' Primary Matters of Concern Regarding Services*

In addition to individual circumstances being the most important value, one can begin to see through the negative subordinate categories what providers could perhaps value. Indeed, we know from working with the employment service providers that they were historically required to meet employment targets set by the government. While theories such as agency theory (Eisenhardt 1989) describe the above conflict of interests, we believe that this type of low level theorizing might be missing much of what could be found by *starting* from this point and digging deeper. We do not believe that a principle-agent problem exists at all times between principle and agent – one could give countless examples where this is rarely an issue, especially when both parties are aimed at a similar goal (cf. Collins and Porras 1994). But even moving beyond the principle-agent problem, both parties above ostensibly are aimed at the same goal: to gain employment. Additionally, the job seekers are concerned with position fit and ability to match position fit with job listings would arguably be the best short and long-term strategy for the provider in order to continue meeting targets. Thus, the risk factor differences between the two parties in agency theory do not necessarily apply. Indeed, we could probably go on endlessly discussing various singular dimensional theories that attempt to explain behaviour through 'this' or 'that' only to always come back to an unexplained tension and never move beyond it. Or, alternatively, we could consider possibilities based on a relational view.

## 2.4　A Relational Discussion of the Findings

Simpson (2014a) reviews the interpersonal relationship literature to determine what may be the key drivers behind human action with respect to everyday interpersonal relationships. A key finding was that many people who find themselves in a relationship that either invalidates their current and future identity, or where the person finds themselves helpless and/or dictated to, or both, will likely end the relationship *if* they have an option to do so. If they do not have that option, they might generally become uncooperative (Lewis 1998) - much like the job seekers who do not show for appointments, as they many times, regarding employment servicing, they feel helpless and are indeed dictated to. Implicit in Simpson (2014a) or Lewis (1998), or explicit in many other interpersonal relationship studies is that when relationships go 'wrong' it is not simply one side or the other doing the wronging – healthy relationships are co-created as are toxic relationships co-destructed, and value formation and consequently future value application in either follows (cf. Echeverri and Skalen 2011). This should be considered in light of *how* service providers interact with job seekers. Furthermore, Kaufman and Stern (1988) illustrate the significance of perceived unfairness leading to retained hostility and how this negatively impacts future interaction. In light of these points, consider the following excerpts from the seekers:

1. "some of them [providers] seem to think that they are better than you because they have a job and you can't find one…they are paid to help you so they should not look down on you"
2. "[they] have put me in a depressed state because of it. They did not care about my future."
3. "because most places don't want to go out of their way to help you get any further training to boost your chances of getting a job, You are just a dollar amount to them."
4. "the provider tends to attend to their own agenda and fulfilling their own requirements and ignoring the needs and situation of the job seeker."
5. "A couple of times at my appointments they seem angry."

Now, consider that these job seekers (according to participants and alluded to by the authors above) will then go to their appointments feeling wronged, possibly hostile, but generally 'low' feeling and appearing. Then, consider the possibility that from the provider's point of view it could be yet another complaining freeloader who just wants income support and is keeping the provider from meeting their target because the job seeker is lazy. One can then imagine a performative loop where tensions spiral out of control (cf. Zimmermann et al. 2013) to the point where some job seekers simply find the interaction unbearable or at the very least a far less desirable option particularly if they feel that it is a waste of time / they are not being listened to; and providers may simply shut down and focus on numbers. Both of these outcomes could lead to the job seeker not showing to their appointments.

## 3   Technology, Valence and Future Research

While the previous section pointed to motivations and relationality as an explanation for the problem of non-attendance at appointments, this section also considers valence and the way that information is communicated via technology as possible mediating solutions. Valence in this context refers to positive or negative emotions that can be tied to specific wording, which is reflected in the primary matters of concern. In other words, texts that use job seekers' positive matters of concern should induce positive valence. This is particularly important with respect to the reminder that is sent to job seekers, as if it induces negative emotions there is a higher chance that the job seeker will not show to their appointment or exacerbate an already negative relationship. While one possible way forward would be to now go test all of the ways that the above relationship is in fact what we theorized as a possibility, we prefer an approach based in pragmatism (Dewey 1929; Goldkuhl 2011) where we will go ahead and assume that the relationship is indeed 'bad' and look for relational ways to mend it rather than continuing indefinitely to define what 'is'. This is due to job seekers letting us know what is most important to them (their primary matters of concern), but the overall discourse in the scripts is that these concerns are not being addressed or met. We propose a multi-phased approach for tackling the issues.

First, technology is the easiest of all the entities to control and is therefore ideally the place to begin. The first step would be to get the seekers to return to an employment provider, and currently the last form of communication the seekers will have before the appointment is the SMS message. Positive valence could be achieved in the message by creating a reminder that is framed from our initial findings (what the job seekers would like for the service providers to acknowledge), for example: "we at [provider x] **understand your personal circumstances**, and **look forward to seeing you** on [datetime] at [address] to **help you find suitable work**". The same could apply to all other forms of electronic communication, such as email, web portal information, etc.

> Proposition #1: SMS messages framed to create positive valence based on the values of seekers will increase overall appointment showings.
>
> Proposition #2: SMS messages framed to create negative valence based on the values of seekers will decrease overall appointment showings.

Prospect Theory (Kahneman and Tversky 1979) offers several informative points with respect to further understanding valence in decision making. The theory states that when an individual is faced with a decision where there are two options – one option offers certainty of gain where the other offers possible loss – individuals tend to choose the option that offers certainty (ibid). However, if there are two options that both represent loss, then the individual tends to choose the *more* risky option, not the less risky one.

　　　　Proposition #3: SMS messages that include a financial certainty option for showing and a financial loss for not showing will increase appointment showings.

　　　　Proposition #4: SMS messages that include two options that both result in financial loss will decrease appointment showings significantly.

Propositions #2 and #4 above raise ethical and social issues that would require great care in their testing, and might only be ethically tested from an analysis of SMS templates already in use by the government agency and/or the employment service providers. However, future research could dig even deeper into valence by considering 'gain' and 'loss' from an identity standpoint. Prospect Theory only considers financial loss – other theories, such as Personal Construct Theory (Kelly 1955), speak to identity loss as the primary source of gain or loss and are in line with the aforementioned interpersonal relationship literature, pragmatism, and relationality. We propose a methodology similar to that of (Simpson 2014b) where the researcher conducts a Repertory Grid interview followed by laddering to core personal constructs in order to understand values that go far beyond what is normally considered and tell us about the person's philosophy on life (cf. Fransella 2003).

　　　　Proposition #5: The gathering and application of core values with respect to providers will further enhance the effectiveness of positively framed SMS messages.

These values, by the nature of the method, are mapped to many other lower level concrete people or technologies that the researcher can use to help the seeker compare and contrast among them – the result of which is a relational understanding of those people and technologies. In other words, we should be able to further understand what makes a 'good' provider by comparing them with non-providers (for example). Additionally, comparing and contrasting various technologies (e.g. SMS, email, web portal, smart phone applications) might give us insight into how various technologies are viewed by the seekers. This might allow for us to begin thinking about how to use technologies to engage job seekers in ways we could not have previously thought (the ways emerge).

Finally, we propose the above methodology to be used to interview the employment service providers as well as the government employees responsible for the providers. Doing so should allow for the mapping of core values among all three, and digital communication among all three can be framed accordingly. This should allow for an increase in multi-level stakeholder alignment (removed for review) and consequently positive valence with respect to communication and interaction. The gathering and application of core values of job seekers, employment service providers, and government managers should increase communication and effective IT use among the groups.

　　　　Proposition #6:　The gathering and application of core values to all forms of digital communication among all stakeholders will increase multi-level stakeholder alignment (outcomes and IT usage).

Furthermore, future qualitative interviews will seek to dive further into the specifics of why job seekers feel as though their needs are not being met (primary matters of concern) in order to iteratively improve future SMS texts.

## 4　Conclusion

This paper answers calls for research into individual level conceptual analyses of value in the eyes of non-US citizens with respect to e-government (Bélanger and Carter 2012). We have provided a relational view surrounding interpersonal and digital communication tensions between government contracted service providers and job seekers. We have also provided a theoretical framing that illustrates how to move beyond simply pointing out and/or testing the principle-agent problem and work towards a solution based on Prospect Theory and Personal Construct Theory, further answering calls for research by Bélanger and Carter (2012) to consider theories outside of the discipline (economics and psychology, respectively). Each of our propositions, including multi-level stakeholder alignment, can indeed be tested, measured, and re-theorized if needed in order to home in on solving what is an important practical problem as well as allowing for new theoretical insights to emerge.

The consequences of ignoring these views are far from trivial and have implications far beyond simple costs and expenditures. The aforementioned sustained hostility can easily lead to further disengagement of citizens. This could perpetuate the views of both sides, which could further agitate all involved, thus preventing the effective delivery of employment services in an accessible, timely, equitable and financially viable manner. Furthermore, the relational effects do not just stop there – this is a highly political topic and therefore has a relationship to the rest of society – one that is now networked, and the exploration of the information frontier entails an exponentially increasing number of encounters with worldview invalidating information (cf. Williams et al. 2012). We believe that this simple example provides an illustration of a phenomenon that each and every one of us faces in everyday life. In a relational world, multi-level stakeholder alignment involves us all, along with all of our interests (and disciplines), everyday, in everything we do.

# 5　References

## Copyright